\documentclass[]{spie}  
 
\usepackage{amsmath,amsfonts,amssymb}
\usepackage{graphicx}
\usepackage[colorlinks=true, allcolors=blue]{hyperref}

\title{Characterization of atmospheric turbulence \\ for the Large Synoptic Survey Telescope}

\author{Claire-Alice H\'ebert}
\author{Bruce Macintosh}
\author{Patricia R. Burchat}
\affil{Kavli Institute for Particle Astrophysics and Cosmology, Stanford University, \\ Stanford, CA 94305, USA}

\authorinfo{Further author information. Send correspondance to C.-A. H\'ebert, e-mail: chebert@stanford.edu}

\begin{document} 
\maketitle

\begin{abstract}
One of the scientific goals of the Large Synoptic Survey Telescope (LSST) is to measure the evolution of dark energy by measuring subtle distortions of galaxy shapes due to weak gravitational lensing caused by the evolving dark matter distribution.
Understanding the point spread function (PSF) for LSST is a crucial step to accurate measurements of weak gravitational lensing. Atmospheric contributions dominate the LSST PSF. Simulations of Kolmogorov turbulence models are commonly used to characterize and correct for these atmospheric effects. 
In order to validate these simulations, we compare the predicted atmospheric behavior to empirical data.
\\
The simulations are carried out in GalSim, an open-source software package for simulating images of astronomical objects and PSFs. Atmospheric simulations are run by generating large phase screens at varying altitude and evolving them over long time scales. We compare the turbulence strength and temporal behavior of atmospheres generated from simulations to those from reconstructed telemetry data from the Gemini Planet Imager (GPI). GPI captures a range of spatial frequencies by sampling the atmosphere with 18-cm subapertures. \\
The LSST weak lensing analysis will measure correlations of galaxy ellipticity, requiring very accurate knowledge of the magnitude and correlations of PSF shape parameters. Following from the first analysis, we use simulations and sequential short exposure observations from the Differential Speckle Survey Instrument (DSSI) to study the behavior of PSF parameters -- e.g., ellipticity and size  --  as a function of exposure time. These studies could help inform discussions of possible variable exposure times for LSST visits — for example, to provide more uniform depth of visits.
\end{abstract}


\section{INTRODUCTION}

Among other scientific goals, the Large Synoptic Survey Telescope (LSST) aims to improve the precision of cosmic shear measurements. 
These measurements are performed by measuring spatial correlations among distortions of galaxy shapes that result from the bending of light as it travels through the mass distribution in the expanding universe.
In order to make an unbiased measurement of cosmic shear, it is crucial to correct for correlated noise present in survey images.
This motivates the need for a robust (i.e., unbiased) model of the LSST point-spread function (PSF), which encapsulates the effects of atmospheric seeing and optics.
Both the magnitude and spatial correlations of the PSF shape parameters must be very accurately known.

The dominant contribution to the single-exposure PSF for the LSST is atmospheric turbulence; as wind blows turbulent patches of air across the telescope aperture, it introduces spatial and temporal correlations in the PSF. 
It is important to model these contributions accurately, and current methods rely on simulations of atmospheric turbulence. 
Here we use data from the Gemini Planet Imager (GPI) to validate these simulations and explore the simulation input parameter space. 
The GPI instrument uses adaptive optics (AO) to successfully image exoplanets at Gemini South, less than 2\,km from the LSST site on the Cerro Pach\'on ridge in Chile. 
The signal from the GPI AO telemetry can be used to reconstruct wavefronts, which can then be compared to the simulation-based predictions.
These data are described in Section\,\ref{sec:gpidata}.

Another important question pertains to the behavior of the atmospheric PSF with varying exposure time.
At short exposures, the PSF exhibits a characteristic speckle pattern, and can be quite elliptical; however, as exposure time increases, these speckles average out to a smoother, more circular PSF. 
The magnitude of the ellipticity and spatial correlations among PSF shape parameters are expected to decrease as this averaging occurs, but how quickly they reach their asymptotic values is unknown. 

This time scale is particularly relevant for the LSST, which currently has a default plan of pointing the telescope at each patch of sky for 30\,s, split into two 15\,s exposures -- much shorter, for example, than the 90\,s exposures for the Dark Energy Survey (DES), or the even longer exposures for Hyper-Suprime Cam (HSC). 
Is this default exposure time for LSST  in the ``long exposure time'' regime?
Moreover, the 30\,s exposure itself is under discussion; for example, LSST could use the full 30\,s for a single exposure, or forgo the fixed 30\,s exposure per visit in favor of variable exposure times. 
The time scales of atmospheric PSF variability can inform this discussion -- in particular, the behavior of PSF shape and ellipticity, which are particular relevant for measurements of cosmic shear.

The impact of exposure time is investigated using data from the Differential Speckle Survey Instrument (DSSI), also mounted on Gemini South. 
These data correspond to series of short-exposure images of stars, which are in essence movies of the PSF. 
We stack images to approximate PSF averaging in real data for different exposure times and use these to test the fidelity of simulations. We extract PSF parameters and compare the real and simulated behavior as a function of integrated exposure time. 
This process is described in Section\,\ref{sec:psf}.

The simulations used here are implemented in GalSim, a software package for simulating astronomical objects and PSFs, and are described in more detail in Section\,\ref{sec:sim}. Results are presented and discussed in Section\,\ref{sec:wfresults} and \ref{sec:psfparams}.

\section{Data and Methods}
\subsection{AO telemetry from GPI}
\label{sec:gpidata}
The GPI adaptive optics system consists of a Shack-Hartmann wavefront sensor and two microelectricalmechanical system (MEMS) deformable mirrors\cite{GPI}. 
Over a 60\,s GPI exposure, the wavefront sensor uses 2$\times$2 quad-cell centroiding to measure the slope of the wavefront on a 48$\times$48 grid of subapertures (18\,cm sampling). 
These measurements and corrections are performed at 1\,kHz speed in the AO control system.

The phase of the wavefront is reconstructed in Fourier modes.
High frequency and low frequency modes are separated and corrected by the tweeter and woofer deformable mirrors, respectively \cite{LisaP}. 
By saving the commands sent to these mirrors, one can later reconstruct the phase of the wavefront at the pupil\cite{Adam}.

The reconstructed wavefront telemetry from GPI can  provide information about the atmospheric conditions and parameters at Cerro Pach\'on. 
In particular, by comparing simulated wavefronts to these data, we can validate certain aspects of atmospheric simulations (e.g., in Galsim). 
In addition, the data can serve to inform the choice of input simulation parameters. 
To explore the effect of various parameters, we calculate the temporal power spectrum of the wavefront, as well as the variance of the phase across the pupil. 
The datasets used here are 20\,s or 30\,s long.
Results are discussed in Section\,\ref{sec:wfresults}.

\subsection{Fast speckle images}
\label{sec:psf}
The DSSI camera is designed to take speckle images simultaneously in two colors \cite{dssi}. 
This is achieved by using a dichroic beamsplitter to split the light in two filters, which are then imaged on two independent electron-multiplying CCDs (EMCCDs).
The data used in the following analysis was taken with a blue/red dichroic and filters with central wavelengths of 692\,nm and 880\,nm, and widths of 40 and 50\,nm, respectively. 
This simultaneous imaging improves signal to noise for astrometric measurements, and provides color information of the stars that DSSI images.
In our context, it enables us to study the chromaticity of atmospheric seeing.

   \begin{figure} [ht]
   \begin{center}
   \begin{tabular}{c} 
   \includegraphics[height=9.5cm]{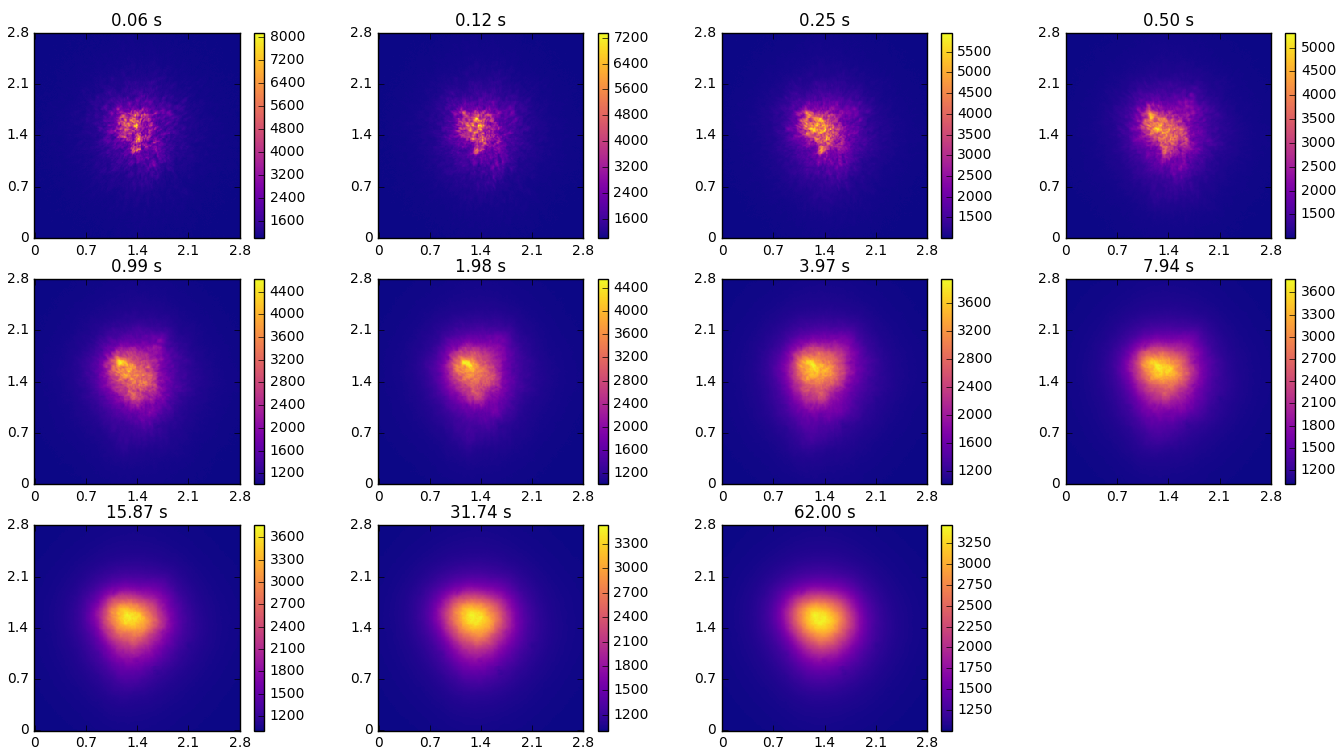}
   \end{tabular}
   \end{center}
   \caption[f1] 
   { \label{fig:psf} Snapshots of the integrated flux in a set of DSSI exposures. Each image has double the number of integrated exposures compared to the preceding image. The time given above each image is the total exposure plus readout time.}
   \end{figure} 

The instrument uses two EMCCDs with a 0.011\,arcsec pixel scale. 
A 256 pixel $\times$ 256 pixel subarray is read out from each CCD, yielding a 2.8\,arcsec field of view. 
Each dataset is composed of 1000 sequential images, each with an exposure time of 60\,ms. 
After each image is taken, the instrument takes roughly 2ms to read out the data. 
The integrated exposure time (i.e., the time over which photons are collected) is thus 60\,ms, and the elapsed time, which includes the 2ms readout, is slightly longer at 62\,ms.
Nine data sets are analyzed here. 
All images were recorded when the star was close to zenith on nights with a range of seeing conditions.

These data are accumulated to form a long-exposure PSF by simply stacking the short exposure images. 
The results are shown in Figure \ref{fig:psf}. 
By eye, at the end of the 60\,s exposure, we see a smooth, if not circular, PSF with no visible speckles. 
The default LSST exposures of 15\,s and 30\,s also appear to be fairly smooth. 

For a quantitative assessment of the PSF, we must extract parameters by fitting a profile to these images.
This is done using a Kolmogorov profile\cite{GalSim}, which is the expected PSF shape for an infinite exposure time, with pure Kolmogorov turbulence. 
This profile is circularly symmetric, so we fit a shear to the profile to make an elliptical PSF. 
The best fit profile parameters are found by minimizing the $\chi^2$ between this elliptical Kolmogorov profile and each PSF image.

\subsection{Simulations}
\label{sec:sim}
The images described in the previous sections are compared with images generated with GalSim\cite{GalSim}, an open-source package for simulating PSFs and astronomical objects. 
The simulation represents the 3D atmosphere as a series of 2D turbulence layers, which are free to translate due to wind as the simulation progresses. 
At each time step in the simulation, the wavefront at the telescope pupil is found by a two process. 
First, sum the contribution of turbulence from each layer (i.e. the projection of the telescope aperture through each layer), and then take the Fourier transform of this sum.

The results discussed here are from simulations of three layers of atmosphere, each of which uses Von K\'arm\'an turbulence \cite{vonK} with an outer scale of L$_0 = 25$\,m. 
Unless specified, the simulations used Taylor's frozen-flow hypothesis \cite{Taylor}. 
The parameters for the turbulence and wind for each layer are given in Table 1. 
The altitude and Fried parameter r$_0$ for each layer are taken from median seeing atmospheric measurements at Cerro Pach\'on\cite{Adam}.
The wind speed for each layer is drawn randomly from a uniform distribution between 0 and the maximum wind speed, which increases with altitude as shown in the table.

For comparison with the AO telemetry data, simulations were generated for continuous integrated exposures of 20 or 30 second, depending on the data of interest. 
The simulation time step used was 1\,ms, to match the cadence of the GPI AO system.
To test dependence of wavefront variation on turbulence strength, a second set of simulations was generated with  Fried parameters a factor of $2^{3/5}$ smaller than  those listed in Table \ref{table:1} (i.e., with  turbulence strength $C_n^2$ a factor of 2 smaller). 
For a third set of simulations, we used the original values of $r_0$ shown in the table, but modified the ``frozen-flow'' hypothesis by introducing atmospheric boiling: every 10 time steps (i.e., every 0.01\,s), a small fraction $1-\alpha$ of the turbulence amplitude in the phase screens is replaced by a random realization of Van K\'arm\'an turbulence. 
The fraction used here is $\alpha = 0.99$, so that $1\%$ of the phase is replaced. 
In each of the three sets of simulations, wind directions were drawn randomly for each turbulence layer.
The simulations were run with different numbers seeding the random generation of wind parameters/turbulence screens. 
 
Speckle simulations were generated to mimic the exposure and readout times during DSSI data collection -- i.e., a series of 1,000 60\,ms exposures separated by 2\,ms of dead time, for a total simulation time of 62 seconds. 
After checking convergence of results, the simulation time step was set to 15\,ms.
The wind directions were determined a bit differently for these simulations; 
instead of randomly drawing a direction for each layer, we first assign a primary direction for the particular simulation and then choose a small random deviation from this direction according to a centered normal distribution with $\sigma=5^\circ$ for each layer.  
To match the nine DSSI data sets used in the analysis, we ran nine simulations with different random seeds.
 
\begin{table}[h!]
\caption{Simulation parameters for the three turbulence layers. 
The wind speed for each layer is drawn randomly from a uniform distribution between 0 and the maximum wind speed.}
\begin{center}
\begin{tabular}{ |c|c|c|c| } 
 \hline
Layer altitude [km] & max wind speed [m/s] & r$_0$ [cm] \\ 
\hline
0 & 10 & 19 \\ 
0.5 & 20 & 43 \\ 
7.6 & 30 & 36 \\ 
 \hline
\end{tabular}
\end{center}
\label{table:1}
\end{table}

\section{Results and Discussion}
\subsection{Characterizing turbulence using wavefronts}
\label{sec:wfresults}
The variance of the reconstructed wavefronts across the telescope pupil provides one measure of the strength of turbulence.
For GPI AO wavefront data and for GalSim simulations,  we compute the variance at each point in time and then histogram the variance for a set of time samples.   
The peak of such a histogram is a measure of the average turbulence strength, and the spread indicates how much the strength varies over the total time -- 20\,s  for this analysis. 
\begin{figure} [ht]
   \begin{center}
   \begin{tabular}{c} 
   \includegraphics[height=7cm]{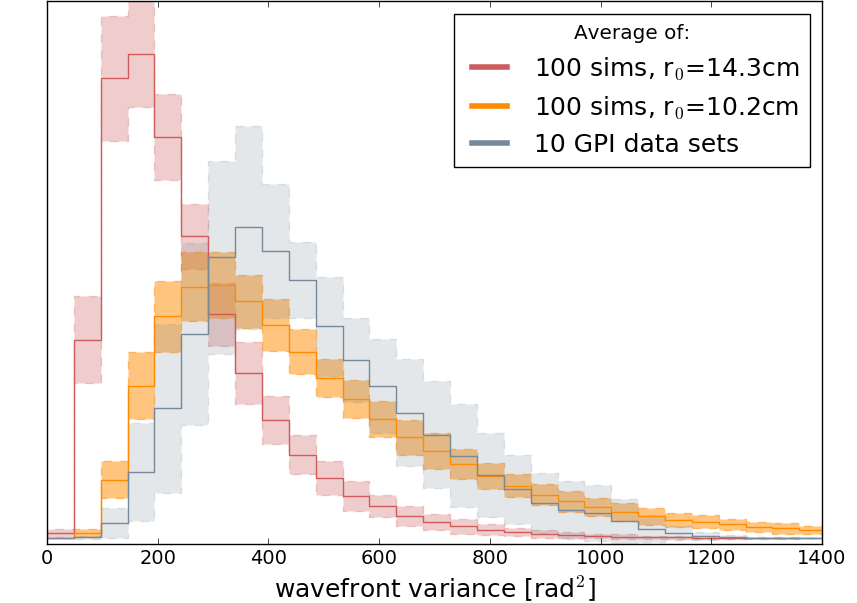}
   \end{tabular}
   \end{center}
   \caption[f1] 
   { \label{fig:var} 
   Histograms of the wavefront variance across the telescope aperture computed for each kHz sample for GPI data and two simulations with different Fried parameters (20,000 variances corresponding to 20\,s total time).
The solid bold histogram corresponds to the average variance in each bin for 10 data sets or 100 simulations while the shaded region indicates the $\pm 1 \sigma$ spread.}
   \end{figure}

Three variance histograms are plotted in Figure \ref{fig:var}: 
for reconstructed GPI wavefront data, and 
for two simulations with turbulence strengths that are different by a factor of two, as described in Section\,\ref{sec:sim}.  
The solid bold histogram of each color corresponds to the average variance in the bin for 10 datasets (gray) or 100 simulations (orange and red), and the shaded regions indicate the $\pm 1\sigma$ (standard deviation) spread in the values of the variance in each bin. 
We see that in the simulations, the distribution of wavefront variance depends strongly on the value of $r_0$, and that the simulation with stronger turbulence (r$_0 \sim 10.2$\,cm) is a much better match to these datasets than (r$_0 \sim 14.3$\,cm). 
Rather than quote the $r_0$ values for each of the layers in the simulation, we report the effective Fried parameter.
For a set of atmospheric layers indexed by i, the effective Fried parameter is given by: $(\sum_{i} r_{0,i}^{-5/3})^{-3/5}$.

While r$_0$ is an important parameter in determining the turbulence strength, we see that in both the simulations (with fixed values of r$_0$) and the data the variance has a wide spread about the peak value.

\begin{figure} [ht]
   \begin{center}
   \begin{tabular}{c} 
   \includegraphics[height=10cm]{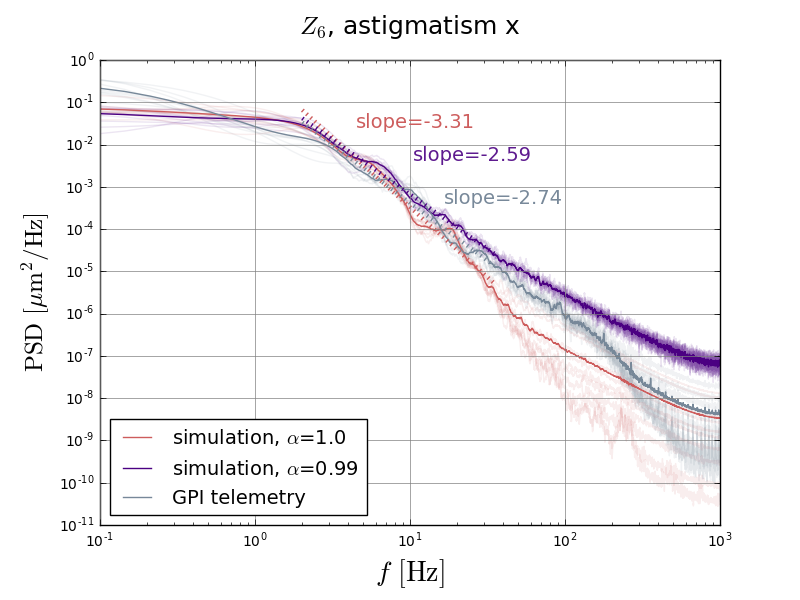}
   \end{tabular}
   \end{center}
   \caption[f2] 
   { \label{fig:psd} Temporal power spectral density for Zernike coefficient $Z_6$ for GPI data (gray) and two sets of simulations (boiling off in red; boiling on in purple). 
The slopes are extracted by fitting a power law to the approximately linear region of the curves between 2\,Hz and 40\,Hz. 
Each pale line is the PSD from one simulation/dataset, and the average of the ten is shown in bold.
}
   \end{figure}

The amount of power at each temporal frequency -- called the power spectral density (PSD) -- is another way of characterizing turbulence\cite{Adam}. 
In particular, we can compare the slopes of the PSDs for data and simulations with the power law slope expected for a Kolmogorov turbulence model. 
As outlined in Ref \citenum{Adam}, we decompose the wavefronts from reconstruction of GPI telemetry and from simulations into Zernike functions, a complete basis set of orthogonal functions describing optical aberrations. 
The PSDs are then computed for each Zernike coefficient.
As an example, we plot in Figure \ref{fig:psd} the PSD for the coefficient corresponding to $Z_6$, as smoothed periodograms. 
The dark and pale curves correspond to the mean and spread, respectively, for ten realizations of two simulations  -- boiling off ($\alpha = 1.0$) and boiling on ($\alpha = 0.99$) -- and ten datasets.
   
The expected slope of the power spectrum for spatial Kolmogorov turbulence is $-\frac{11}{3}\sim -3.67$. 
We expect the temporal power spectrum to have this same value due to the correspondence between spatial and temporal evolution of turbulence screens. 
We compare the Kolmogorov slope with those calculated from the PSDs in Figure \ref{fig:psd}, for data and for the two simulations.
We fit the slopes from the middle region of the spectrum (2 to 40 Hz) where the curves are approximately linear. 
In this region, the simulation with no atmospheric boiling is the closest to Kolmogorov, but its slope is less steep than the $-\frac{11}{3}$ prediction -- perhaps due the presence of an outer scale, or the finite nature of the 20-second interval. 
Either of these factors could suppress power at low frequencies, reducing the overall slope. 
The GPI data slope has an even weaker dependence on frequency than this first simulation. 
However, we find that adding atmospheric boiling to the simulation results in better agreement with the data. Varying the boiling parameter $\alpha$ could lead to a closer match, as could  varying other parameters that influence the PSD slope -- e.g. the outer scale. 

In the next section, we present results on the temporal evolution of atmospheric PSFs. The study of spatial characteristics of wavefronts in this section could be used to inform choices of parameters in those simulations.

\subsection{Evolution of PSF parameters}
\label{sec:psfparams}

\begin{figure} [ht]
   \begin{center}
   \begin{tabular}{c} 
   \includegraphics[height=8cm]{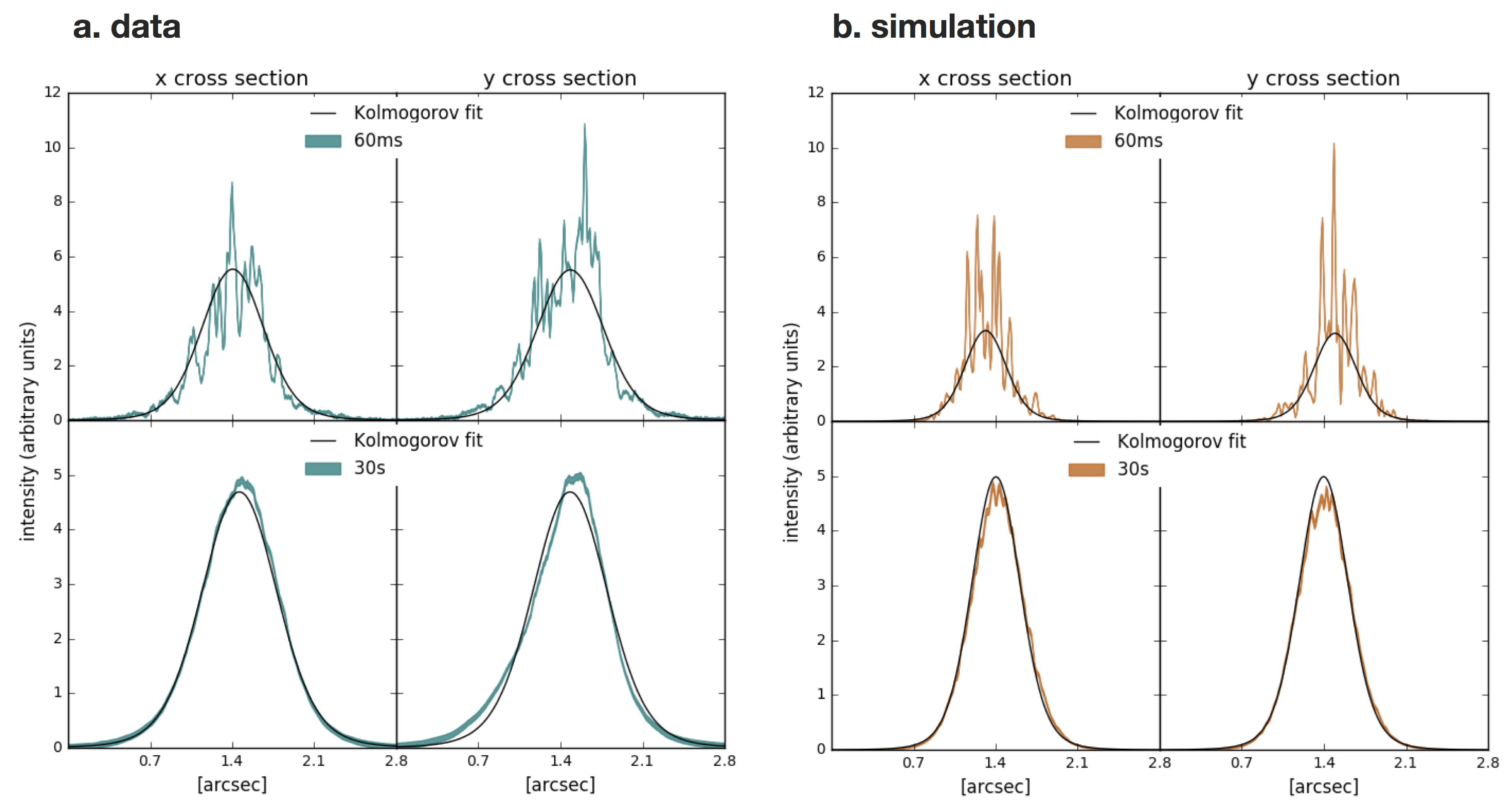}
   \end{tabular}
   \end{center}
   \caption[f2] 
   { \label{fig:cross} Cross sections through the centroid of the PSF for (a) DSSI data images and (b) simulated PSFs. 
   Top panels correspond to a single 60\,ms exposure, and bottom panels to a 30\,s integrated exposure.    Left and right columns correspond to x and y cross sections, respectively, for the pixel values (colored curves) and best fit Kolmogorov models (black curves).
}
\end{figure}

As described in Section\,\ref{sec:psf}, short exposure speckle images from the DSSI camera are stacked to form an accumulating PSF (Figure \ref{fig:psf}). 
These are fit to elliptical Kolmogorov profiles;
sample fits are displayed in Figure \ref{fig:cross}, where we plot cross sections of both PSF images and models. 
Each panel shows a cross section -- i.e., the pixel values -- through the centroid of the image (determined from the fit to the PSF), as well as the cross section of the best fit model for that image. 
For both data (Figure \ref{fig:cross}a) and simulation (Figure \ref{fig:cross}b), we show results for a single 60\,ms exposure (top panels) as well as for the nominal LSST exposure time of 30\,s  (bottom panels). 
We note that the spikes in the 60\,ms PSFs are not noise, but rather the speckles expected in these short exposures. 
We see that these almost completely disappear after 30\,s of integration, in data and simulations.

We find that the Kolmogorov model captures much of the PSF shape in the DSSI image and in the simulated PSF once the speckles begin to average out, consistent with the long-exposure definition of the Kolmogorov profile.
In the y cross section of the long-exposure DSSI image (lower right panel of \ref{fig:cross}a.), we see the data is asymmetric compared to the fit so that the long-exposure image does not converge exactly to a Kolmogorov model.
This is less obvious in the x cross section.

A priori we do not expect the Kolmogorov profile to correctly model the speckles present in the short-exposure images, and therefore fit parameters may not be meaningful for these time scales.
For example, the fits to the short-exposure simulated PSFs appear to systematically underestimate the pixel flux in the images, which leads to underestimated PSF size (half-light radius) for the speckle-dominated images.
Nevertheless, we use the PSF parameters extracted from the fits to study the temporal behavior of the atmospheric PSF, keeping in mind that parameters extracted for short exposure times may not be meaningful. 

\begin{figure} [ht]
   \begin{center}
   \begin{tabular}{c} 
   \includegraphics[height=8cm]{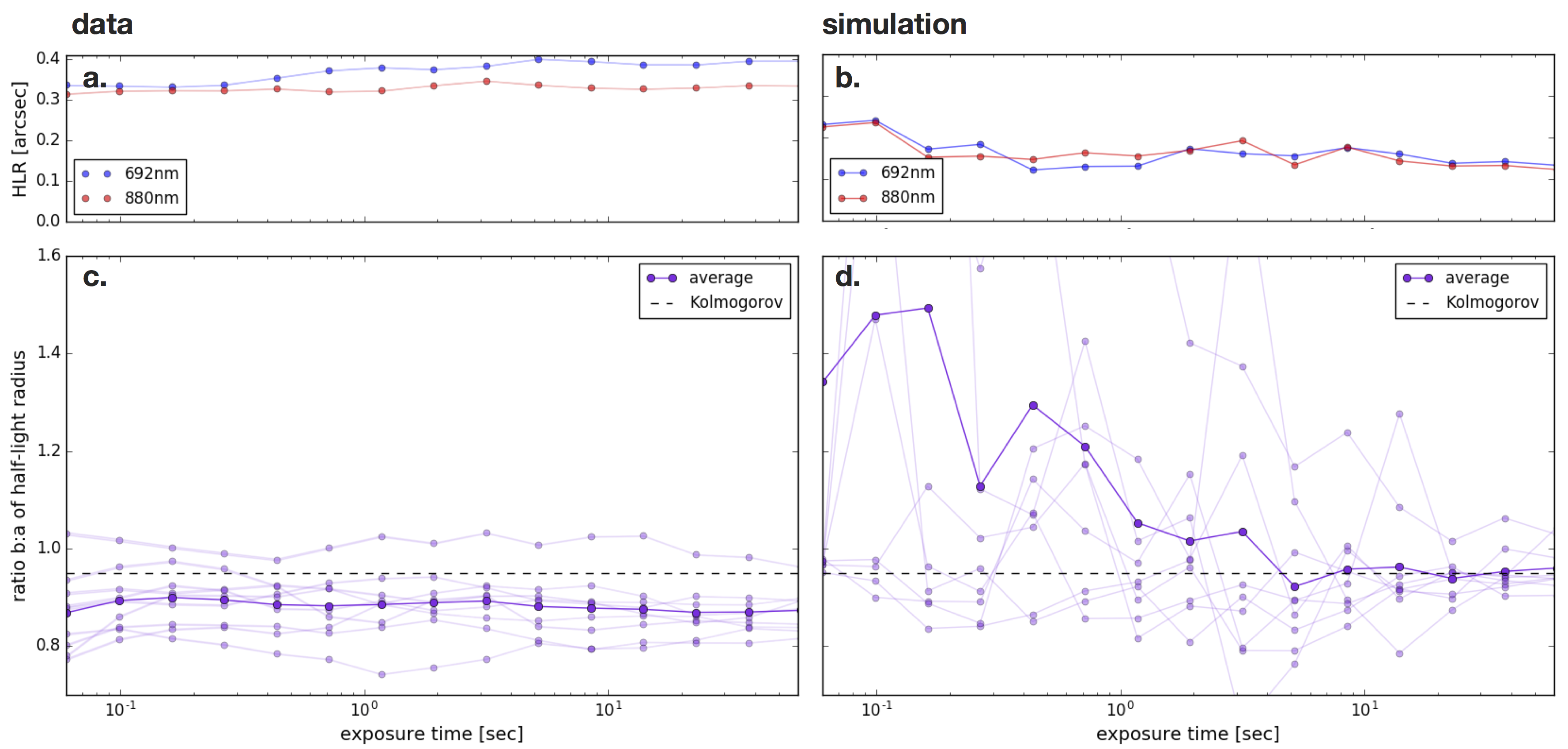}
   \end{tabular}
   \end{center}
   \caption[f2] 
   { \label{fig:hlr} Top panels:  Half-light radius (${\rm HLR}$) as a function of exposure time for (a) a  single DSSI dataset and (b) a simulated dataset, in two filters. 
   Bottom panels: Ratio of half-light radii in two filters, ${\rm HLR}(880\,{\rm nm})/{\rm HLR}(692\,{\rm nm})$, for (c) DSSI data and (d) simulation. 
   The dark and pale curves correspond to the mean and spread, respectively, for nine DSSI datasets and nine simulated PSFs.
   The dashed lines in the lower panels correspond to the ratio of PSF sizes predicted by Kolmogorov turbulence evaluated for the mean wavelengths of the filters; see Eq.\,\ref{eq:kol_color}.
}
   \end{figure}

An important parameter in discussing PSFs for almost any application, including cosmic shear, is its size.
Here we parametrize size with the half-light radius (${\rm HLR}$). 
In panels a and b of Figure \ref{fig:hlr}, we plot the ${\rm HLR}$ (extracted from the Kolmogorov fits described above) for one example of DSSI data images and for a simulated PSF, as a function of integrated exposure time, from 60\,ms to 60\,s.  
The sizes for data and simulation behave rather differently with exposure time, with the data PSF size increasing with exposure time while the simulated PSF size decreases; 
this is especially noticeable for short exposure times. 
The expected behavior is for the PSF size to increase with exposure time since the centroid of the PSF moves around with time.
The behavior of the simulation may be due to the fit  underestimating the size of the PSF in the presence of speckles as discussed earlier. 

In Figure\,\ref{fig:hlr}a and b, we also show the ${\rm HLR}$ for two filters for DSSI data and for two wavelengths  for the simulation ($\lambda_a$=692\,nm, $\lambda_b$=880\,nm), enabling an analysis of the color dependence of the PSF.
We expect the PSF to be smaller for longer wavelengths than for shorter wavelengths, and indeed the data exhibits this dependence. 
The simulation also converges to this after $\sim10$\,s, but the sizes fluctuate significantly for shorter integration times. 

The bottoms panels of Figure \ref{fig:hlr} show the wavelength-dependent behavior for each of the nine DSSI data sets (c) and for nine simulated PSFs (d). 
Each line corresponds to the ratio of PSF size in the two filters (${\rm HLR}(880\,{\rm nm})/{\rm HLR}(692\,{\rm nm})$) for one dataset; the bold line shows the average ratio. 
As we might expect from the discussion of the single datasets in the top panels, the results for data are much smoother than for simulation.
Only one dataset shows a ${\rm HLR}$ ratio greater than 1, indicating a  PSF larger at longer wavelength, whereas the ratios for the simulations are mostly greater than 1 and exhibit significant fluctuations for exposure times less than a few seconds.

The dashed line in each of these panels shows the expected ratio for a Kolmogorov model:
\begin{equation}
\label{eq:kol_color}
\frac{{\rm HLR}(880\,{\rm nm})}{{\rm HLR}(692\,{\rm nm})} \sim \left(\frac{880}{692}\right)^{-1/5} \sim 0.95\,.
\end{equation}
The data exhibit a stronger wavelength dependence than this prediction, evidence again of non-Kolmogorov behavior, or of a potential systematic effect in the DSSI camera or readout. 
The simulations converge to the Kolmogorov prediction after an exposure time of about 15 seconds.
   
   \begin{figure} [ht]
   \begin{center}
   \begin{tabular}{c} 
   \includegraphics[height=10cm]{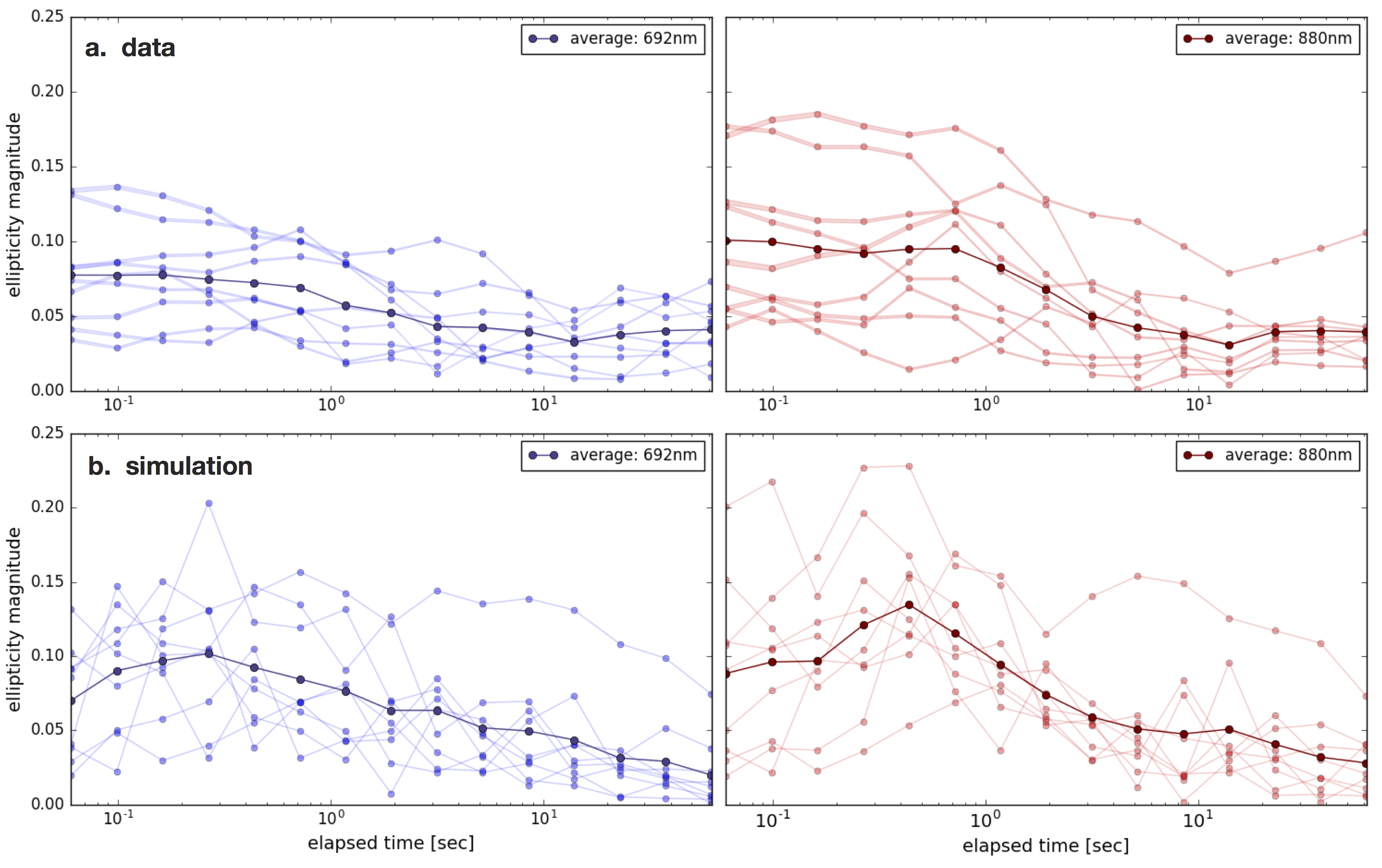}
   \end{tabular}
   \end{center}
   \caption[f2] 
   { \label{fig:g_mag} Magnitude of PSF ellipticity as a function of exposure time. Best fit ellipticities from the integrated exposure in DSSI data (top panels) and simulation (bottom panels). The columns show results for the blue ($\lambda_a=692$\,nm) and red ($\lambda_b=880$\,nm) filters.
}
   \end{figure} 

Another important PSF parameter to consider for potential cosmic shear systematic biases is  ellipticity. 
Figure \ref{fig:g_mag} shows the magnitude of PSF ellipticity in each filters for DSSI data (a) and for the simulation (b). 
As in the previous figure, each pale line is an individual dataset, and the bold lines show the behavior of the mean. 
We do not expect significant evidence of chromatic behavior for ellipticity.
Just as we saw with the ${\rm HLR}$ results, the ellipticity measured for the data is more stable than for simulation. 
Focusing instead on the average behavior, the main difference between the data and simulation is the asymptotic behavior -- the ellipticity in data plateaus at  $\approx$ 15\,s exposure time, while the simulation appear to continue to decrease even at the end of the 60\,s integration time. 

\begin{figure} [ht]
   \begin{center}
   \begin{tabular}{c} 
   \includegraphics[height=10.2cm]{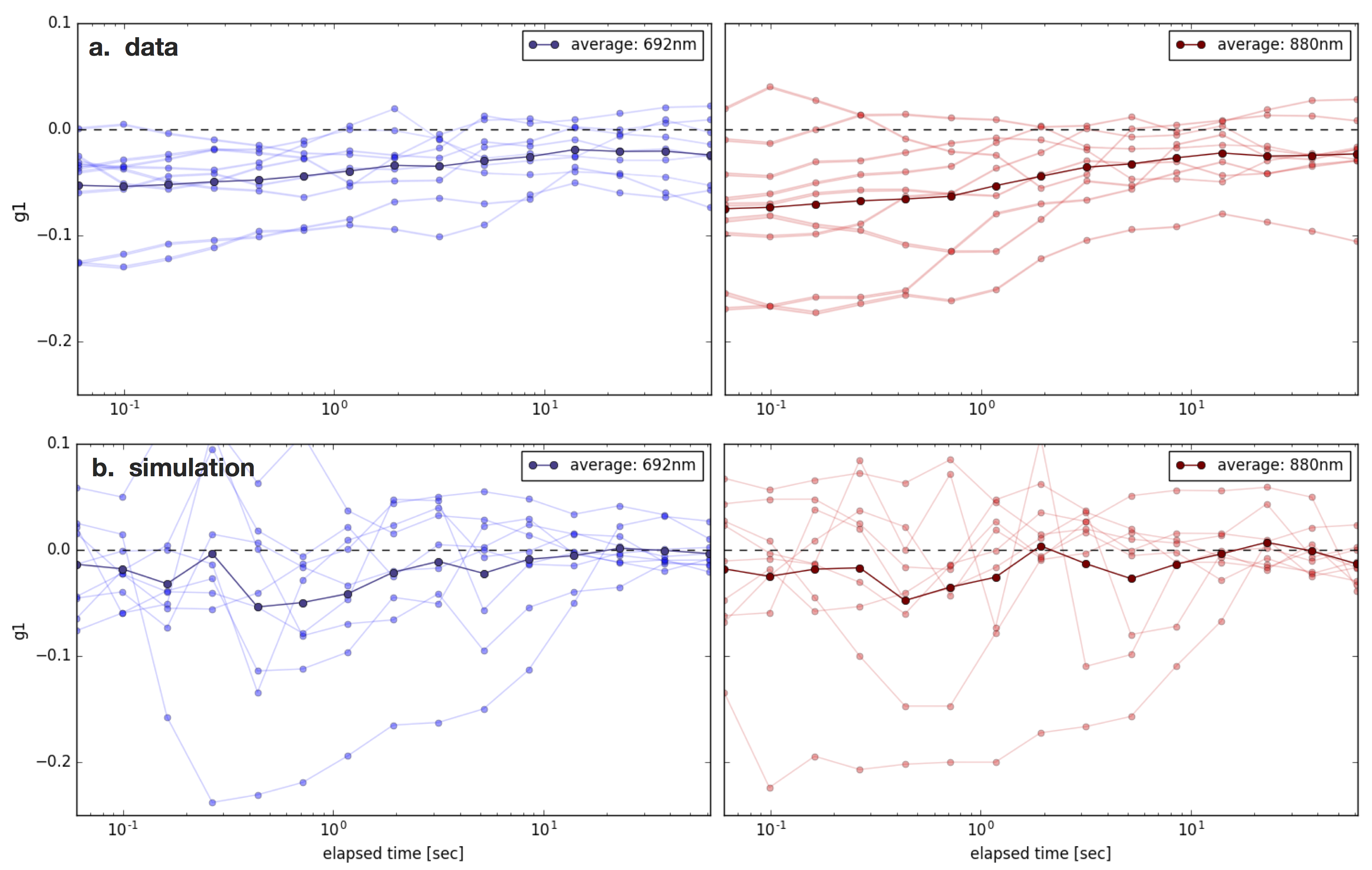}
   \end{tabular}
   \end{center}
   \caption[f2] 
   {\label{fig:g1} Similar to Figure\,\ref{fig:g_mag}, but for the g$_1$ component of ellipticity instead of magnitude. DSSI data in the top panels; simulation in the bottom panels. 
}
   \end{figure}

Ellipticity can be parametrized in terms of two components, g$_1$ and g$_2$, where g$_1$ corresponds to stretch ($>0$) or compression ($<0$) along the x direction, and g$_2$ to the distortions in a direction 45$^\circ$ to the x axis. 
Plotting each of these components individually could give insight into the deviation in asymptotic behavior between the simulation and data -- for example, directional systematic biases related to the rows and columns of pixels in the DSSI sensors or simulation. 
Figure \ref{fig:g1} shows results for g$_1$.
Similar to the ellipticity magnitude, we see that the mean value of g$_1$ over nine data sets has not reached  0 in data in 60 seconds. 
Results for g$_2$ are not shown, but do not have this behavior: the data average does not stray far from zero at any point in the exposure. 
In simulation, this g$_1$ component oscillates and is not as stable as data, but does not show the same overall negative trend. 

One hypothesis is that a prevailing wind might result in this type of persistent ellipticity direction. 
However, simulations using a prevailing wind direction, rather than a random direction for each realization, does not show this effect.
A prevailing wind could move turbulence more quickly along a particular direction, but whether this would lead to more elliptical PSFs is not obvious.  
Alternate hypotheses are that the DSSI optics, sensors or electronics lead to a PSF that is broader in the y direction than the x direction.

\section{Future work}
This study provides a framework for further investigation of atmospheric PSFs. 
The non-vanishing PSF ellipticity in data requires further study.  Is it an atmospheric effect, or an optical or instrumental effect? 
Data are available to systematically analyze more stars -- for example, stars of different magnitudes, with different seeing and different airmass.

A significant concern in the comparison of data and simulation is the lack of knowledge about correct input parameters for the atmospheric simulations --  e.g., wind speed,  wind direction, variation with altitude, etc. 
GPI telemetry from a wider range of seeing conditions could potentially guide this choice, especially when combined with data from seeing monitors and wind information from the National Oceanic and Atmospheric Administration (NOAA) databases.
In particular, these data could inform a realistic range of simulation inputs to improve the study of PSF parameter behavior. 
In general, a greater range of parameters in both observational datasets and simulations will enable a more systematic study of the simulation input parameters and the asymptotic PSF behavior.

\acknowledgments 
C.-A. H\'ebert is supported by the DOE Computational Science Graduate Fellowship (CSGF) program. 
The authors thank E.\,Horch and M.\,Everett (DSSI) for providing the speckle camera data, and the GPI survey team for the AO telemetry.
In addition, we thank Joshua Meyers for his advice on the simulation of atmospheric PSFs in Galsim, and 
Adam Snyder and Aaron Roodman for valuable discussions and insights, and for sharing code for extracting Zernike coefficients.

\bibliography{report} 
\bibliographystyle{spiebib} 

\end{document}